\documentclass[10pt]{iopart}
\usepackage{iopams}
\usepackage{setstack}
\usepackage[dvips]{graphicx}
\usepackage{epsfig}

\begin{document}

\title{Energy Transport in an Ising Disordered Model}

\author{E. Agliari$^1$, M. Casartelli$^{1,2}$ and A. Vezzani$^{1,3}$}
\address{$^1$ Dipartimento di Fisica, Universit\`a di Parma, Viale
G.P. Usberti n.7/A (Parco Area delle Scienze) 43100 - Parma - ITALY}
\address{$^2$ INFN, gruppo collegato di Parma, Viale G.P. Usberti
n.7/A (Parco Area delle Scienze) 43100 - Parma - ITALY}
\address{$^3$
CNR-INFM, National Research Center S3, via Campi 213/a, 41100 -
Modena- ITALY}

\begin{abstract}
We introduce a new microcanonical dynamics for a large class of Ising systems isolated or maintained out of equilibrium by contact with thermostats at different temperatures. Such a dynamics is very general and can be used in a wide range of situations, including disordered and topologically inhomogenous systems. Focusing on the two-dimensional ferromagnetic case, we show that the equilibrium temperature is naturally defined, and it can be consistently extended as a local temperature when far from equilibrium. This holds for homogeneous as well as for disordered systems. In particular, we will consider a system characterized by ferromagnetic random couplings $J_{ij} \in [ 1 - \epsilon, 1 + \epsilon ]$. We show that the dynamics relaxes to steady states, and that heat transport can be described on the average by means of a Fourier equation. The presence of disorder reduces the conductivity, the effect being especially appreciable for low temperatures. We finally discuss a possible singular behaviour arising for small disorder, i.e. in the limit $\epsilon \to 0$.
\newline
keywords: Transport processes (Theory), Heat conduction, Disordered systems (Theory)
\end{abstract}


%

\section{Introduction}
\label{intro}

Microcanonical dynamics seems to be a proper mode to describe an
isolated system, or its isolated bulk, without any assumption on
the equilibrium state between the system and the surrounding. A
typical context where such a dynamics occurs is in the study of
transport properties in continuous or discrete models (see
\cite{livi} for general references). For instance, for a system in
thermal contact with two heat reservoirs at different
temperatures, energy may be exchanged only through the
thermostats, and the internal dynamics requires an energy
preserving evolution to be implemented in the microcanonical
framework; analogous considerations hold for matter transport and
similar items \cite{deutsch,gaspard,baldovin}.

Presently, we are interested in heat transport in discrete models, namely spin systems. As it is well known, for
such systems there exist several rules to simulate microcanonical evolution, e.g. the Q2R rule introduced by Vichniac, the Creutz rule and, more recently, the KQ rule \cite{vich,pom,toff,cmv,acv}. All of them present some weak aspects which require the introduction of some constraints concerning geometrical support, boundary conditions (i.e. the allowed range of imposed temperatures), connectivity, homogeneity, etc. Our purpose is to introduce a microcanonical dynamics sufficiently flexible to be adapted to all these cases. The recent interest in systems displaying quenched disorder and inhomogeneous structure, provides indeed a strong motivation for the development of such a general dynamics \cite{hin,queiroz,dhar,kara,ben}.

As it is well known, Q2R and Creutz dynamics differ in the fact that the latter supplies each site in
the lattice with a possible bounded reserve of ``kinetic'' energy, in such a way that the spin flip
is allowed not only when the magnetic energy is preserved (as in Q2R), but also when the energy excess or defect
may be assigned to or extracted from the reserve, so that  the {\sl total} energy is preserved. This
additional possibility, however, does not solve the dynamical freezing at low energy densities (occurring of course also in Q2R), since in such conditions one easily falls into short periodical cycles, inhibiting the transport of heat and the exploration of the whole set of configurations at constant energy.
In other terms, both Creutz and Q2R dynamics have an ergodicity breakdown at low energy, loosing  the independence of initial conditions which is necessary to ensure a statistical effectiveness  to the dynamical description. Moreover, one should introduce {\sl ad hoc} modifications when the connectivity is not homogeneous.
As to the KQ dynamics, introduced and discussed in \cite{cmv,acv}, even if it possesses the wanted ergodicity, it is not a satisfying solution to the general dynamical problem, since it is specifically meant for regular lattices.

The new dynamics we are going to define actually modifies the Creutz scheme, only maintaining the distinction between magnetic and kinetic energies. As we shall see, not only it displays effective ergodicity for all practical purposes, but it can also be easily adapted to a wide class of spin systems without any substantial modification, exhibiting therefore the desired versatility. In particular, it can be implemented on arbitrary topologies, even non regular ones. Another aspect of this flexibility is that the extension to inhomogeneous systems, where the spin coupling is not a constant, is quite natural. Hence, a great variety of disordered models become easily accessible in principle; in the following we study an example of quenched disorder.

It is worth underlining that, in all the cases considered here, the magnetic and kinetic energies result to be non correlated: this allows a natural definition of temperature at equilibrium, which, in a very natural way, depends only on the average kinetic energy. This approach can be consistently extended, when far from equilibrium, to a definition of {\sl local} temperature. Now, dealing with heat transport, this implies an important improvement in the description of steady states, since not only diffusivity but also conductivity become easily computable. In particular, in studying a model of quenched disorder, the interesting point is to explore how inhomogeneity affects the thermal and transport properties. For instance, we verify that the conductivity becomes smaller as the disorder grows. The emergence of non analytical behaviour in the limit of vanishing disorder is also observed.

In the following, we first describe in details how our microcanonical dynamics works (Section \ref{sec:dinamica}) and we show that it is able to relax the system to well defined steady states (Section \ref{sec:checks}), even in the presence of disorder (Section \ref{sec:disordine}), verifying that a temperature as an intensive quantity independent of
the actual values of couplings can still be defined. Then, we focus on the transport properties (Section \ref{sec:transport}), and on the effect of disorder on the conductivity (Section \ref{sec:effect}).
The last section is devoted to our conclusions and final remarks (Section \ref{sec:conclusions}).

\section{Microcanonical Dynamics} \label{sec:dinamica}

Let us consider an Ising model defined on a generic network. In particular, $s_i=\pm 1$ denotes the $i$-site spin; on each link $i \sim j $ we introduce a local exchange interaction $J_{ij}$, so that the magnetic energy  of a link is $J_{ij} s_i s_j$. Moreover, on the link $i\sim j$ we define a local {\it kinetic energy} $ E_{ij}>0$. We remark that while standard Creutz scheme specifies a bounded amount of ``kinetic'' energy in each site, we introduce an energy defined on links which is  non-negative but, in principle, unbounded. The dynamical rule proceeds as follows:
\begin{enumerate}
\item Start from a distribution of energies $E_{ij}$;
\item choose randomly a link  $i \sim j$;
\item extract one over the possible four spin-configurations for the couple of sites $i,j$, and evaluate the magnetic energy variation $\Delta E^m$ induced by the move, where $\Delta E^m$ allows for energy variations occurred on the link  $i \sim j$ as well as on those pertaining to links adjacent to sites $i$ or $j$;
\item if $\Delta E^m \leq 0$ accept the move and increase the link energy $E_{ij}$ of $\Delta E^m$. When $\Delta E^m >0$, accept the move and decrease the link energy of $\Delta E^m$ only if $E_{ij} \geq \Delta E^m$.
\end{enumerate}
Three main points have to be remarked:
\begin{itemize}
\item Every link is defined by its  extremes not only in rectangular lattices but in all graphs, even in the case of variable connectivity. This ensures a great flexibility with respect to the underlying geometry.
\item  The interchange between sites and links  allows the energy to travel throughout the graph, even at low energy density, without the freezing of the standard Creutz-Q2R rules or the geometrical restrictions of the KQ dynamics.
\item
The constraint of uniform couplings $J_{ij} \equiv J = const$, typical of previous microcanonical dynamics, can also be relaxed as our dynamics allows to use local (i.e. non uniform) spin interactions.
\end{itemize}
Clearly, the above dynamics conserves the total energy given by the following {\it Hamiltonian} function
\begin{equation}\label{ham}
H(s_i,E_{ij})=\sum_{i \sim j } \left( J_{ij} s_i s_j + E_{ij}\right),
\end{equation}
where the sum runs over all the links of the network. Notice that energy is locally conserved, namely long-range energy transfers are forbidden and hence we can define our dynamics as microcanonical, where the kinetic energy energy $E_{ij}$ works as an additional degree of freedom.
Expression (\ref{ham}) is very general, representing a conserved energy for a model defined on a generic network. Moreover, according to the choice of the couplings $J_{ij}$, a wide class of models (disordered interactions, vacancies, spin glasses etc.) can be recovered.

Actually, for comparisons and checks, we consider first the homogeneous system  $J_{ij} \equiv J$ (with $J \equiv 1$ in simulations) on the standard rectangular lattice of $L_X$ columns and $L_Y$ rows, both in the toroidal geometry (periodical boundaries) and in the cylindrical geometry with two opposite boundaries. Such boundaries can be ``free'' (i.e. three neighbours for each site) or ``open'' (i.e. in contact with a reservoir). Heat conduction in the latter case was studied e.g. in \cite{grant,saito} using Q2R-Creutz dynamics, in \cite{cmv,acv} using KQ dynamics and in \cite{amal} using a layer heat-bath.

Afterward, we shall study  the transport properties in a cylindrical system where the couplings $\{   J_{ij} \}$ are chosen randomly in the interval $[1-\epsilon,1+\epsilon]$, i.e. a disordered ferromagnet.

\section{Temperature and Dynamical checks} \label{sec:checks}

At equilibrium, i.e. in a closed system where no energy is injected or subtracted, every move is reversible, and the probability $P(S\to S')$ to  go from the configuration $S$ to the configuration $S'$ equals $P(S'\to S)$. Detailed balance is satisfied, and the equilibrium probability distribution is constant over all configurations.

If all the configurations with the same energy are visited in the evolution, the system is ergodic and its equilibrium properties can also be described by Boltzmann statistics.
In this perspective, we check the agreement between the numerical stationary results and the Boltzmann distribution, focusing first on the homogeneous system $J_{ij} \equiv J$ defined on a two-dimensional torus; a positive response to the check has to be seen as an indication of effective ergodicity.

A first step in this direction consists in  testing the decorrelation between the kinetic energy $E_{ij}$ and the  magnetic energy $J s_i s_j $ of a link. Indeed, we verify that for each link
\begin{equation} \label{decorr}
\langle E_{ij} J s_i s_j \rangle  =  \langle E_{ij} \rangle \langle J s_i s_j \rangle ~,
\end{equation}
where $\langle \cdot \rangle$ denotes a time average along our dynamics. Kinetic and magnetic part of the Hamiltonian can therefore be treated separately.

\begin{figure}
\resizebox{54mm}{60mm}{\includegraphics{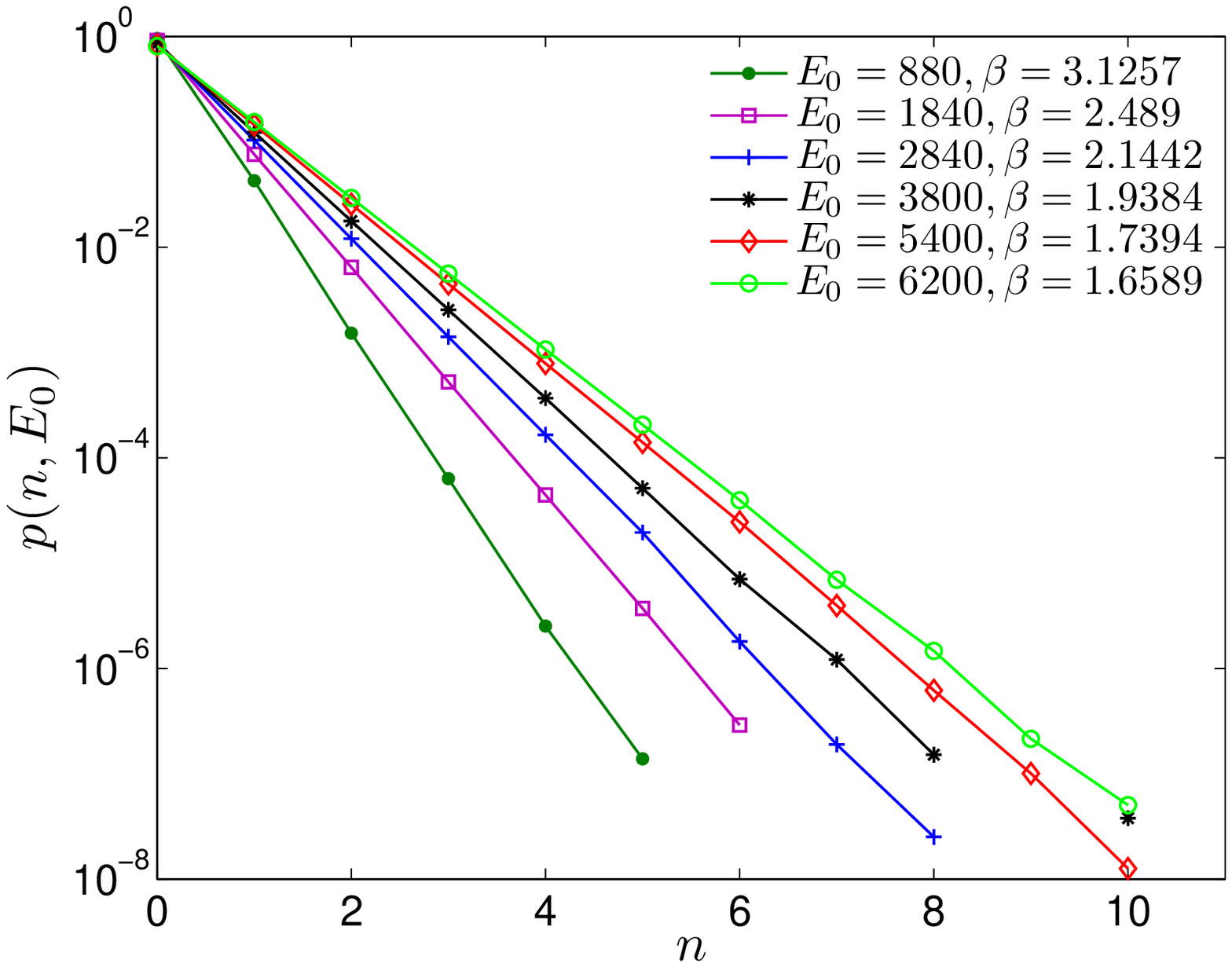}}
\resizebox{54mm}{60mm}{\includegraphics{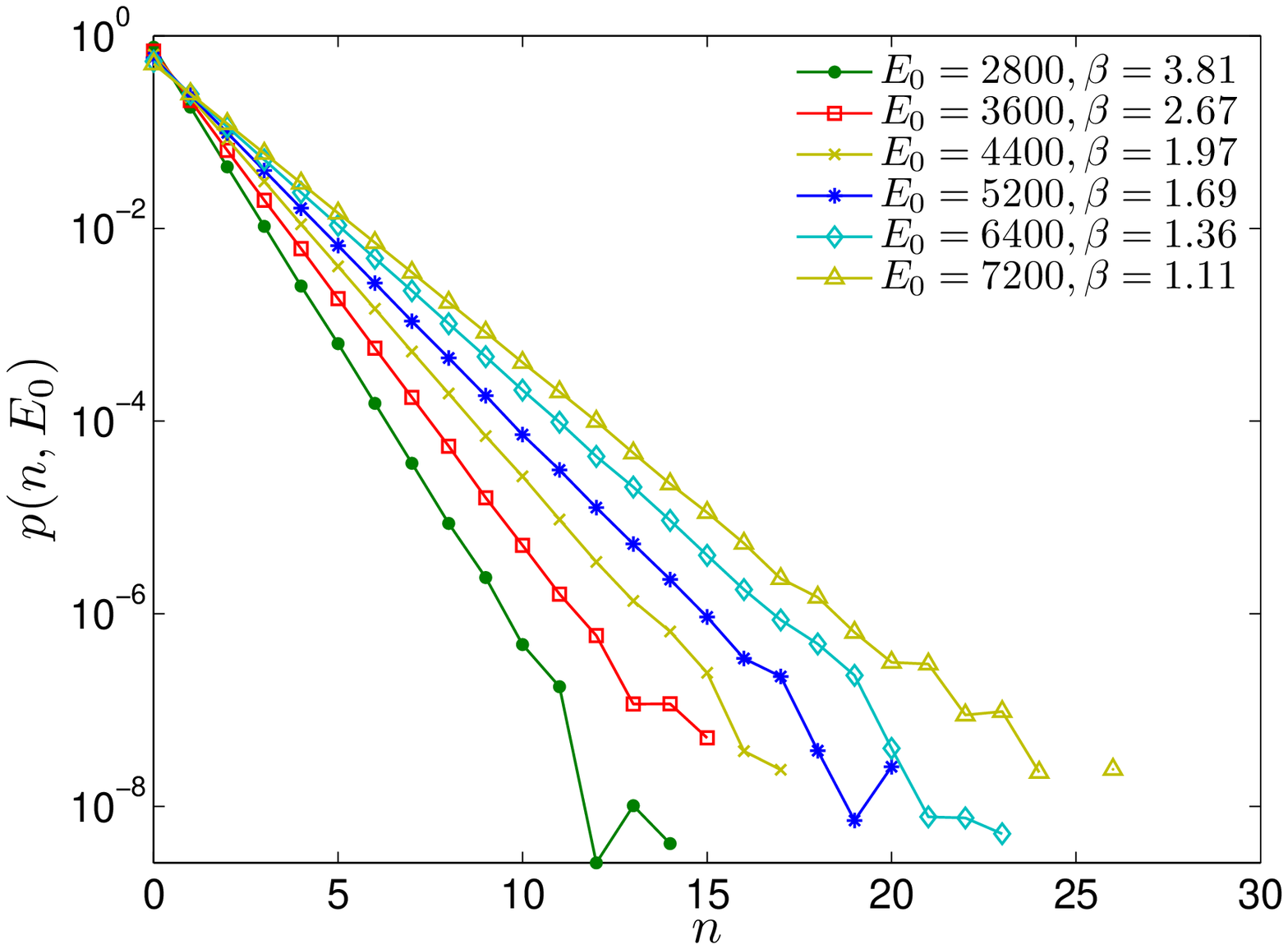}}
\caption{Probability distribution of the kinetic energy per link for an Ising system defined on a square torus of size $L_X=L_Y=48$ with constant coupling strength $J$ (left panel) and with random $J$ and disorder $\epsilon=0.2$ (right panel). Results found for different initializations $E_0$ are reported as shown by the legend. Notice that in the right panel the continuous energy distribution $p(E,E_0), E \in \mathbb{R}$ is approximated by the histogram $p(n,E_0), n \in \mathbb{N}$.}
\label{Distributions}
\end{figure}

As a second point, we observe  that the link energy satisfies the Boltzmann distribution $\exp(-\beta E_{ij})$,  where the fitted constant $\beta=1/T$ is naturally the inverse temperature of the system (Fig.~\ref{Distributions}, left-panel). In general, for the homogeneous system we can write $E_{ij}=4J n_{ij} + \xi_{ij}$, where $n_{ij}$ is an integer and $\xi_{ij}$ is a constant depending on the initial conditions; since $\xi_{ij}$ does not have any influence, hereafter we set $\xi_{ij}=0$. Notice that the factor $4$ multiplying $J n_{ij}$ is due to the fact that the minimal amount of exchanged energy is $4J$ ($E_{ij}$ derives from $\Delta E^m$ and, apart from possible borders, in rectangular lattices the coordination number is even).

The equilibrium temperature has been verified to be independent from the initial distribution of energies and from the particular link considered, as expected from an intensive quantity. Averages on the links are possibly  used only to accelerate the numerical convergence.

The simplest way to evaluate the temperature from the above distribution is to relate $\beta$ with the average link-energy, obtaining
\begin{equation} \label{emdis}
\langle E \rangle ~ =~  { {\sum_{n=0}^{\infty} 4Jn e^{-4J\beta n} } \over  {\sum_{n=0}^{\infty} e^{-4J\beta n} } }
= { 4J \over {e^{4\beta J}-1 }}
\end{equation}
which may be easily inverted to get $\beta $:
\begin{equation} \label{emdis2}
\beta  =  \frac{1}{4J} \log \left( 1+ \frac{4J}{ \langle E \rangle } \right).
\end{equation}

\begin{figure}
\resizebox{0.9\columnwidth}{!}{
\includegraphics{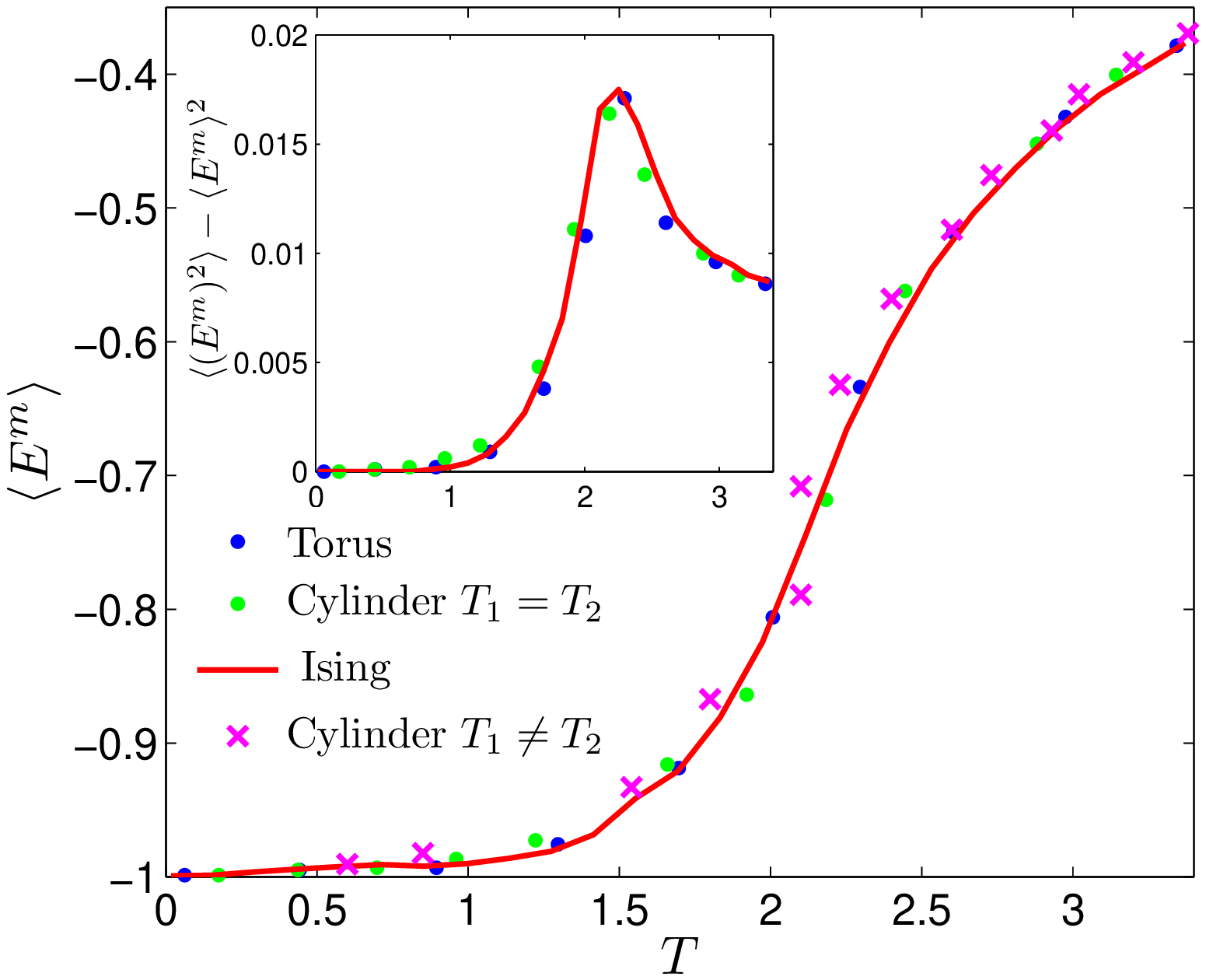}}
\caption{Average magnetic energy per spin $\langle E^m \rangle$ (main figure) and its fluctuations $\langle (E^m)^2 \rangle - \langle E^m \rangle^2$ (inset) as a function of the temperature $T$ for Ising systems of size $N=2304$ defined on a torus and analyzed with the Metropolis canonical dynamics (continuous line) and with the microcanonical dynamics introduced here (dark $\bullet$); the case of an Ising system defined on a cylinder endowed with two thermostats set at the same temperature (bright $\bullet$) or at different temperatures ($\times$) is also considered. The consistency among all cases is clear.}
\label{Cilindro}
\end{figure}
In a subsequent check, we verify that $T =1/\beta$ given by formula (\ref{emdis2}) is the temperature even for the magnetic variables of the system. Indeed, Fig.~\ref{Cilindro} evidences that the average magnetic energy $\langle E^m \rangle$ and its fluctuations $\langle (E^m)^2 \rangle - \langle E^m \rangle^2$ are the same in the microcanonical system at temperature $T$ given by Eq.~\ref{emdis2} and in a two-dimensional Ising model in equilibrium at temperature $T$.

Equivalent results are also obtained introducing thermostats at the boundaries of the microcanonical system (Fig.\ref{Cilindro}). The coupling between system and thermostats is imposed by extracting the link energies at the borders every time-step according to the probability distribution $ e^{-4 \beta J n}$. We remark that in presence of such thermostats the whole system equilibrates at temperature $1 /\beta$, independently of the initial conditions, proving the reliability of such a simple model for thermostats.

Energy transport is naturally introduced by imposing different temperatures at the boundaries. Then,
whenever a stationary state is reached, Equation  (\ref{emdis2}) provides a simple definition of local temperature. Clearly, along the same column all energies and temperatures are expected to be equal, as they are in fact. The consistency of this  definition of local non-equilibrium temperature has been further verified by showing that, at the same  $T$, the local magnetic energy for a link is the same at equilibrium or in the presence of transport (Fig. \ref{Cilindro}).

In conclusion, the homogeneous system follows a quasi equilibrium picture
consistent with the results given by the KQ dynamics \cite{acv}. The remarkable difference is that, in the present case, the existence of a local kinetic energy allows a natural definition of temperature, without any reference to the equilibrium magnetic model.

\section{From Homogeneous to Disorderd Lattices} \label{sec:disordine}

After testing our dynamics for the Ising model on a regular lattice with homogeneous coupling, we exploit the fact that, as anticipated, this dynamics is naturally defined even in more complex situations, e.g. when the underlying topology or the interaction couplings $J_{ij}$ are not homogeneous. Here in particular we focus on the case of a 2-dimensional rectangular lattice where the coupling $J_{ij}$ are randomly chosen from a uniform distribution in an interval, i.e. $J_{ij}\in [1-\epsilon,1+\epsilon]$, $0 < \epsilon <1$; hence, $\epsilon$ is an index of disorder, being the average magnetic interaction fixed and equal to $1$.

\begin{figure}
\resizebox{0.9\columnwidth}{!}{\includegraphics{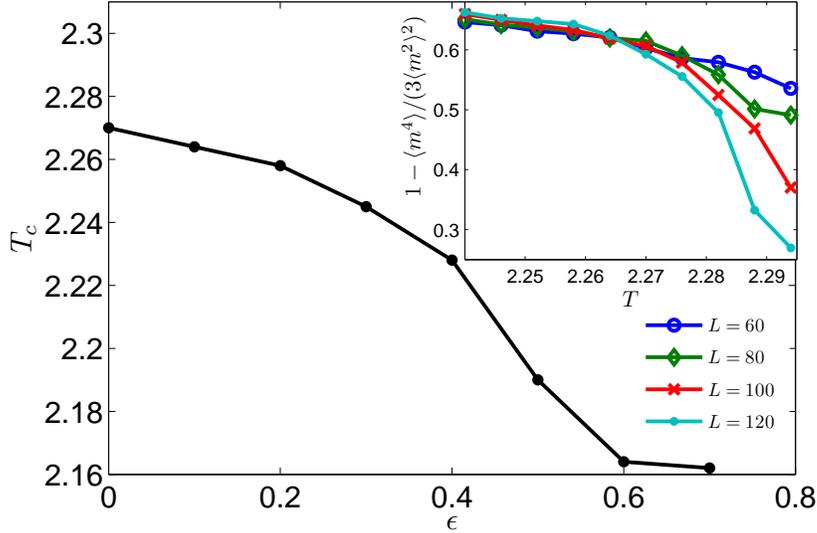}}
\caption{Measure of the critical temperature $T_c$ for systems characterized by different disorder $\epsilon$. The estimate of the critical temperatures has been accomplished by means of Binder cumulants as shown in the inset for the case $\epsilon=0.1$.}
\label{Binder}
\end{figure}

First of all, we study the differences at equilibrium between homogeneous and disordered systems. To this purpose, using the standard Metropolis dynamics, we evaluate the critical temperature $T_C$ by means of Binder cumulant techniques \cite{binder}. Figure (\ref{Binder}) shows the value of the critical temperature as a function of the disorder $\epsilon$. We note that $T_C$ is slightly lowered by the presence of disorder.

Let us now consider the new dynamics for the disordered system. With respect to the homogeneous case $J_{ij}=1$,  the main difference is that the link energy is not of the form $E_{ij}=4 n_{ij}$ with $n_{ij}$ integer. Indeed, the kinetic energy on the link $i \sim j$ can be written as  $E_{ij}= \sum'_{ k\sim h } J_{kh} n_{kh}$, where $E_{ij}\geq 0$, $n_{kh}$ are integers and the sum $\sum'_{k \sim h}$ runs over $i\sim j$ and its six neighbouring links. Hence, in the sum there are seven different real numbers $J_{hk}$ randomly chosen in $[1-\epsilon,1+\epsilon]$ and, if all the combinations of integers  $n_{kh}$ are realized in the dynamical evolution, the values of $E_{ij}$ can be considered densely distributed along the positive real axis (see Appendix).

Therefore, under the hypothesis of ergodicity we expect that the probability density for link energies  is $\exp(- \beta E)$, with $E$ a real number.
In Figure (\ref{Distributions}) we check indeed that in a random system the link energies are distributed according to a continuous exponential function. Also in this case, as in (\ref{emdis}), the temperature can be related to the average kinetic energy, in particular for a continuous variable we have
\begin{equation} \label{emcon}
 \langle E \rangle = {{\int_0^{\infty} z~e^{-\beta z}~dz}\over { \int_0^{\infty} e^{-\beta z}~dz} } = {1 \over \beta}.
\end{equation}
Once again, the constant $1/\beta$ is  the temperature $T$ and the link (kinetic) energy equals the system temperature. We check that, although the system is inhomogeneous, the temperature is the same for all the sites, as expected at equilibrium, and also that $\langle E \rangle$ is a consistent definition of temperature even for the magnetic system, as analogously verified for the homogeneous system (see Fig.~\ref{Cilindro}). Therefore, the temperature is the parameter governing the equilibration of the system. In particular, at equilibrium different parts of an inhomogeneous system may display different behaviours but $T$ is constant in the whole sample. For example, if a part of the system is characterized by constant coupling $J_{ij}=J$ while in the rest $J_{ij}$ is chosen randomly, the local kinetic energy is described by discrete and continuous distributions in the different regions respectively. However, the temperature is the same in both. Analogously in a free cylinder (isolated borders) with constant coupling $J$, we have that the border sites  exchange energy with an odd number of neighbours. The minimal amount of exchanged energy  is therefore $2J$ instead of $4J$, giving  a different discrete distribution for  sites in the bulk and at the border. However, we have numerically verified that the temperature does not change.

Thus, in general, for systems with random-incommensurate couplings, Equation (\ref{emcon}) provides a good definition of temperature.
This definition can naturally be extended to the case when different temperatures are imposed at the boundary thermostats, i.e. in the presence of gradients  ($T_2-T_1>0$) and heat transport, as described in the next Section.

\section{Transport, Conductivity, Diffusivity} \label{sec:transport}
As previously explained, our system satisfies the Boltzmann distribution $\exp{(- E/T)}$, so that in order to fix a temperature $\bar{T}$ on a given set of links, it is sufficient to extract the relevant link energies according to the distribution $\exp{(-E/\bar{T})}$, and this is sufficient to establish a consistent coupling between the system and the thermostats. In the following we consider a cylindrical Ising system with boundaries coupled to two thermostats at temperatures $T_1$ and $T_2$, respectively, and we study the transport properties.
First of all, we verify that in the steady state the kinetic energy $\langle E_{ij} \rangle$ is still well defined on each link. However, in such a  non-equilibrium situation, it varies from link to link; hence, hereafter we use $\langle E_{ij} \rangle$ as a definition of \textit{local} temperature $T_{ij}$. We also notice that once the distance from the thermostats is fixed, its spatial fluctuations increases with $\epsilon$. Hence, the dependence displayed by $T_{ij}$ on the magnetic coupling $J_{ij}$ is very small for $\epsilon$ not too large.
Now, in the presence of temperature fluctuations also transport is expected to display non trivial spatial patterns arising from the interplay between non-equilibrium and disorder. As a first step in studying such a complex phenomenology we investigate how disorder affects the average transport properties. In particular
we introduce the column temperature $T_j$ and magnetic energy $E^m_j$ obtained by spatially averaging over all $i$'s $T_{ij}$ and $\langle E^m_{ij} \rangle$, respectively. The emerging cylindrical symmetry makes $j$ the only gradient variable, and in this situation the simplest model for heat transport is provided by  the one-dimensional Fourier equation, where, as usual,
the column index $j$ will be treated as a continuous variable $x$:
\begin{equation}\label{eq:diff1}
\frac{\partial E^{tot}}{\partial t} = \frac{\partial}{\partial x} \left( \kappa \frac{\partial T}{\partial x} \right).
\end{equation}
Notice that $E^{tot}$ is the specific total energy in column $j$, namely $E^{tot}_j=T_j+E^m_j$, and that $\kappa$ is the conductivity which is expected to depend only on the local temperature and on the system parameters ($J$ and $\epsilon$ in this case). Indeed, in the following we verify that energy transport can be studied by means of equation (\ref{eq:diff1}), and we also evaluate $\kappa$ as a function of $T$, evidencing the effect of disorder.

\begin{figure}
\resizebox{0.9\columnwidth}{!}{
\includegraphics{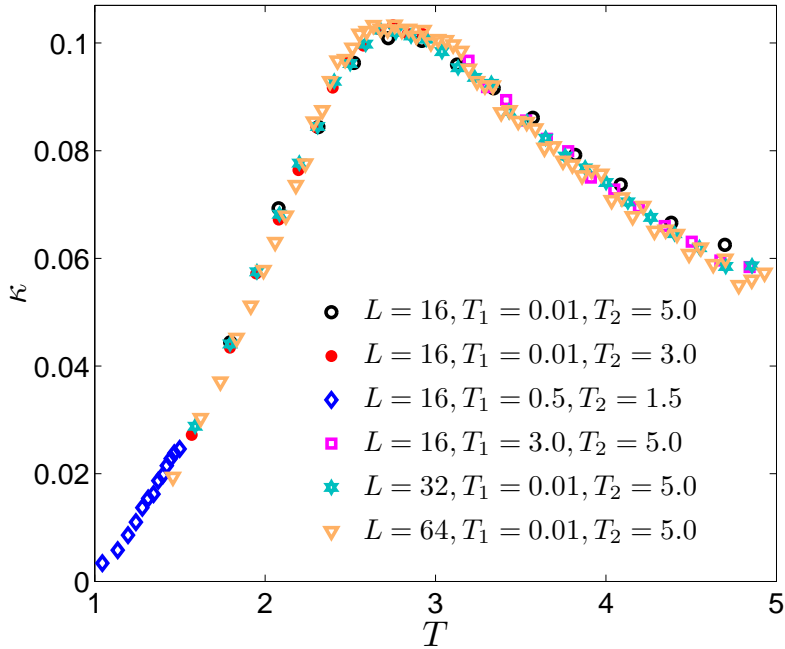}}
\caption{Conductivity versus temperature for an ordered (constant $J$) Ising system. Different sizes $L$ and different temperature gaps $\Delta T$ are considered, as shown in the legend.}
\label{Fig_4}
\end{figure}
\begin{figure}
\resizebox{0.9\columnwidth}{!}{
\includegraphics{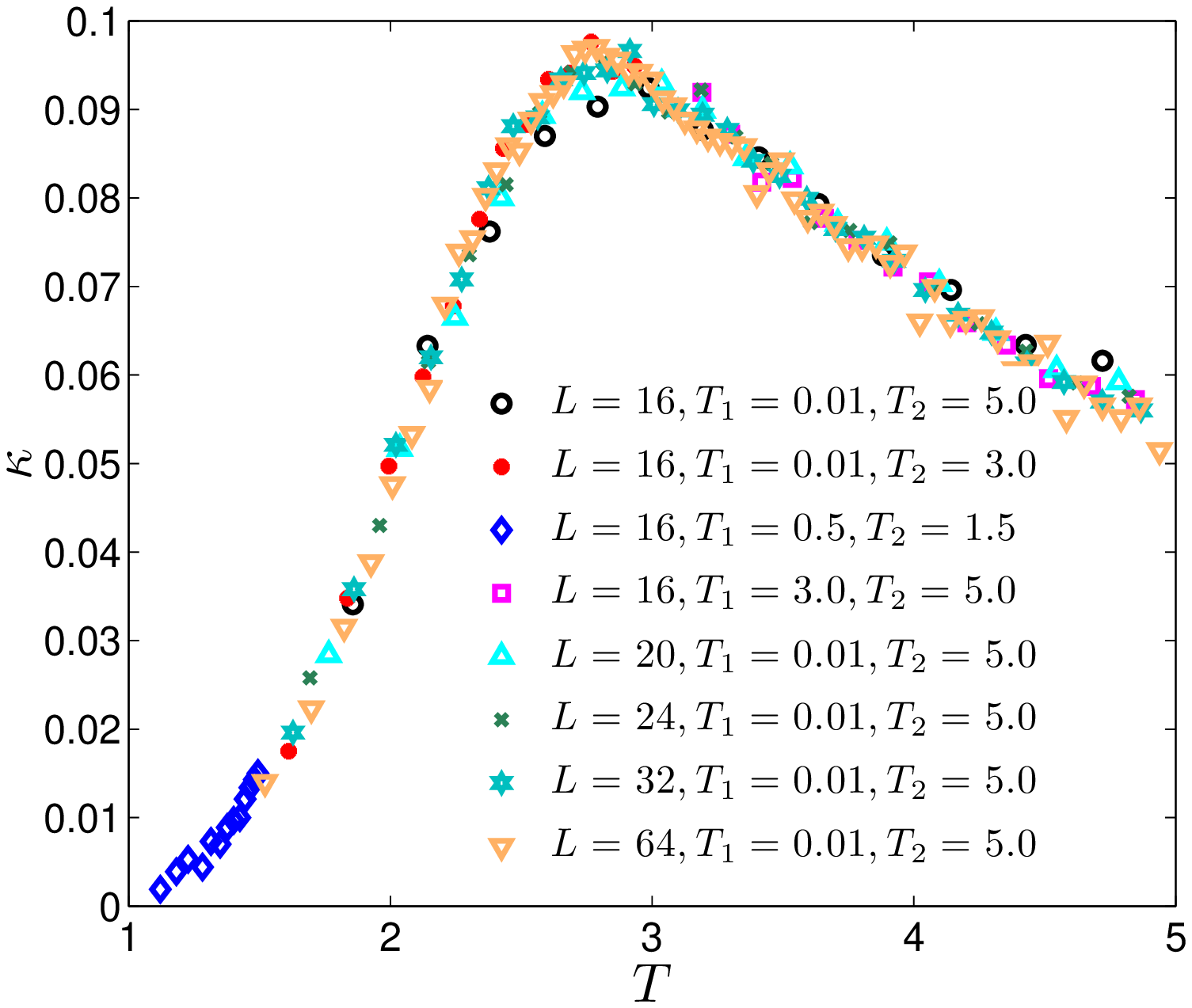}}
\caption{Conductivity versus temperature for a disordered ($\epsilon=0.2$) Ising system. Different sizes $L$ and different temperature gaps $\Delta T$ are considered, as shown in the legend.}
\label{Fig_5}
\end{figure}

In the stationary state, when $\partial E^{tot} / \partial t =0$, we call $Q$ the total amount of heat crossing any given surface, so that $\partial Q / \partial t$ represents the heat flux which can be measured, for instance, from the energy absorbed per unit of time by the thermostat. As shown in \cite{cmv}, we have
\begin{equation}\label{eq:diff3}
\kappa= \frac{\partial Q / \partial t} {\partial T / \partial x};
\end{equation}
clearly $\kappa$  represents the proportionality constant between
the temperature gradient and the flowing heat energy. Similarly,
the diffusivity of the magnetic energy can be defined as:
\begin{equation}\label{eq:diff4}
D=\frac{\partial Q / \partial t} {\partial E^m / \partial x}.
\end{equation}
It is worth underlining that while $D$ can be obtained directly
from the magnetic energy $E^m$ (see Eq.~\ref{eq:diff4}) for the
conductivity $\kappa$ we need a measure of the temperature $T$
(see Eq.~\ref{eq:diff3}). For this reason KQ dynamics, dealing
exclusively with the magnetic energy, allows for the computation
of diffusivity, while here we are able to easily estimate both
quantities.

Introducing the magnetic specific heat $C=\partial E^m / \partial T$,
a comparison of Eq. ~\ref{eq:diff3}  and Eq.~\ref{eq:diff4} provides the simple relation
\begin{equation}\label{eq:D_K}
D=\frac{\kappa}{C}.
\end{equation}
Notice that $C$ is an equilibrium quantity well known for the Ising model.

We have measured conductivity for the ordered system with $\epsilon=0$ (Fig.~\ref{Fig_4}) as well as for the disordered system with fixed $\epsilon \neq 0$ (Fig.~\ref{Fig_5}). In both cases we found a very good collapse of data pertaining to different temperatures intervals and sizes.

\begin{figure}
\resizebox{0.9\columnwidth}{!}{
\includegraphics{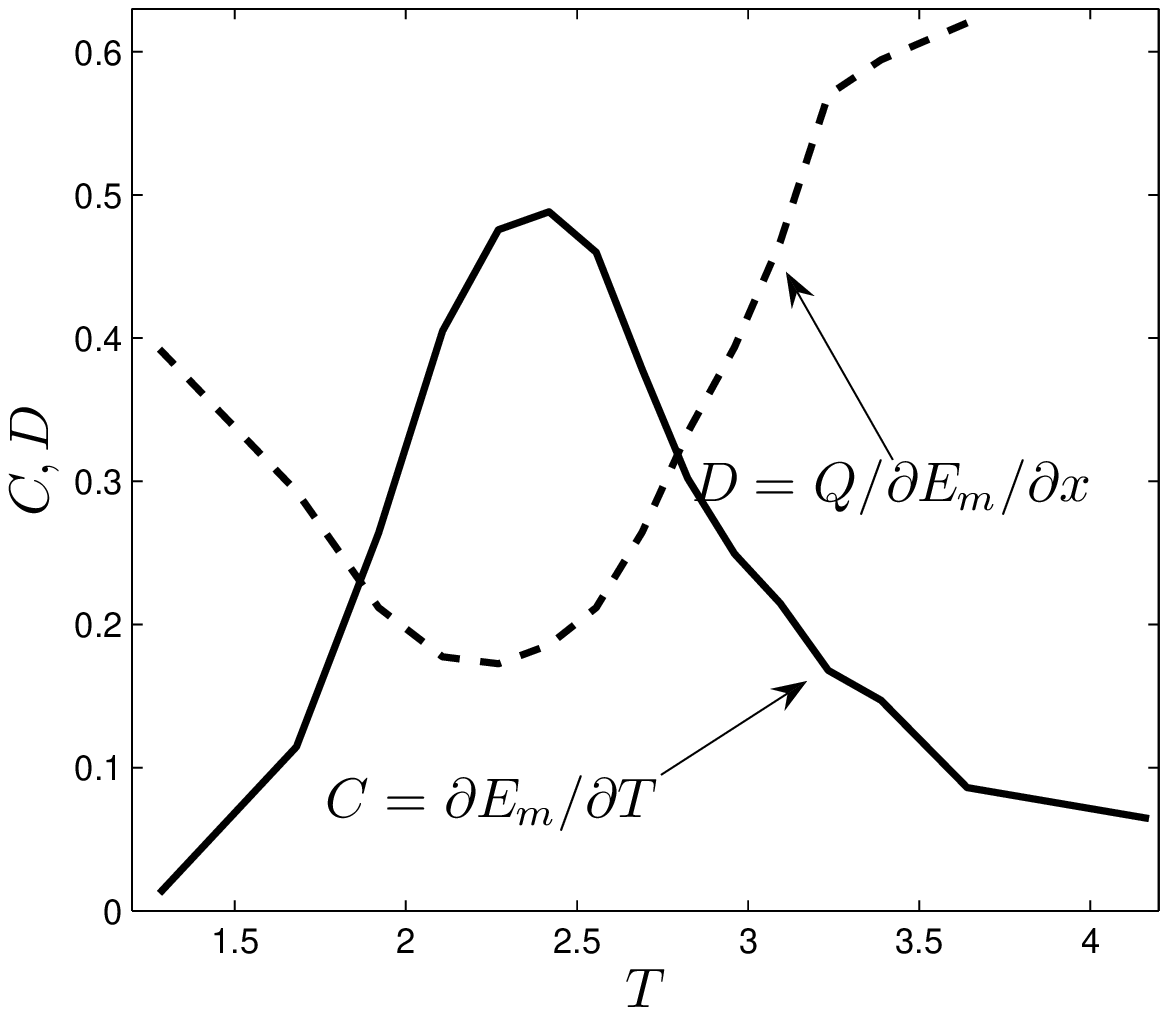}}
\caption{Specific heat $C$ (continuous line) and diffusivity $D$ (dashed line) as a function of the temperature $T$, for an ordered Ising system ($\epsilon=0$) of size $L=16$ and thermostats at temperatures $T_1=0.01, T_2=5.0$.}
\label{Fig_6}
\end{figure}

This provides a strong confirmation of the picture above based on a Fourier transport equation with a temperature (energy) dependent conductivity (diffusivity).
In particular, we notice that the curve for the conductivity exhibits a maximum at a temperature $\tilde{T}$ slightly larger than the critical temperature $T_C$ which corresponds to a peak in the specific heat and to a minimum for the diffusivity (see Fig.~\ref{Fig_6}). As observed also in \cite{acv}, we expect that in the thermodynamic limit the diffusivity goes to zero logarithmically, consistently with the logarithmic divergence of the specific heat, so that the conductivity is not expected to diverge at the critical point $T_C$. Interestingly, from both Fig.~\ref{Fig_4} and Fig.~\ref{Fig_5}, we notice that as the temperature is increased, the rate of growth (for $T<\tilde{T}$) and of decrease (for $T>\tilde{T}$) displayed by the conductivity are sensitively different. This point can be understood by looking at Fig.~\ref{Fig_6} and recalling Eq.~\ref{eq:D_K}: at small temperatures the diffusivity remains finite, while the specific heat is close to zero.

\begin{figure}
\resizebox{0.9\columnwidth}{!}{
\includegraphics{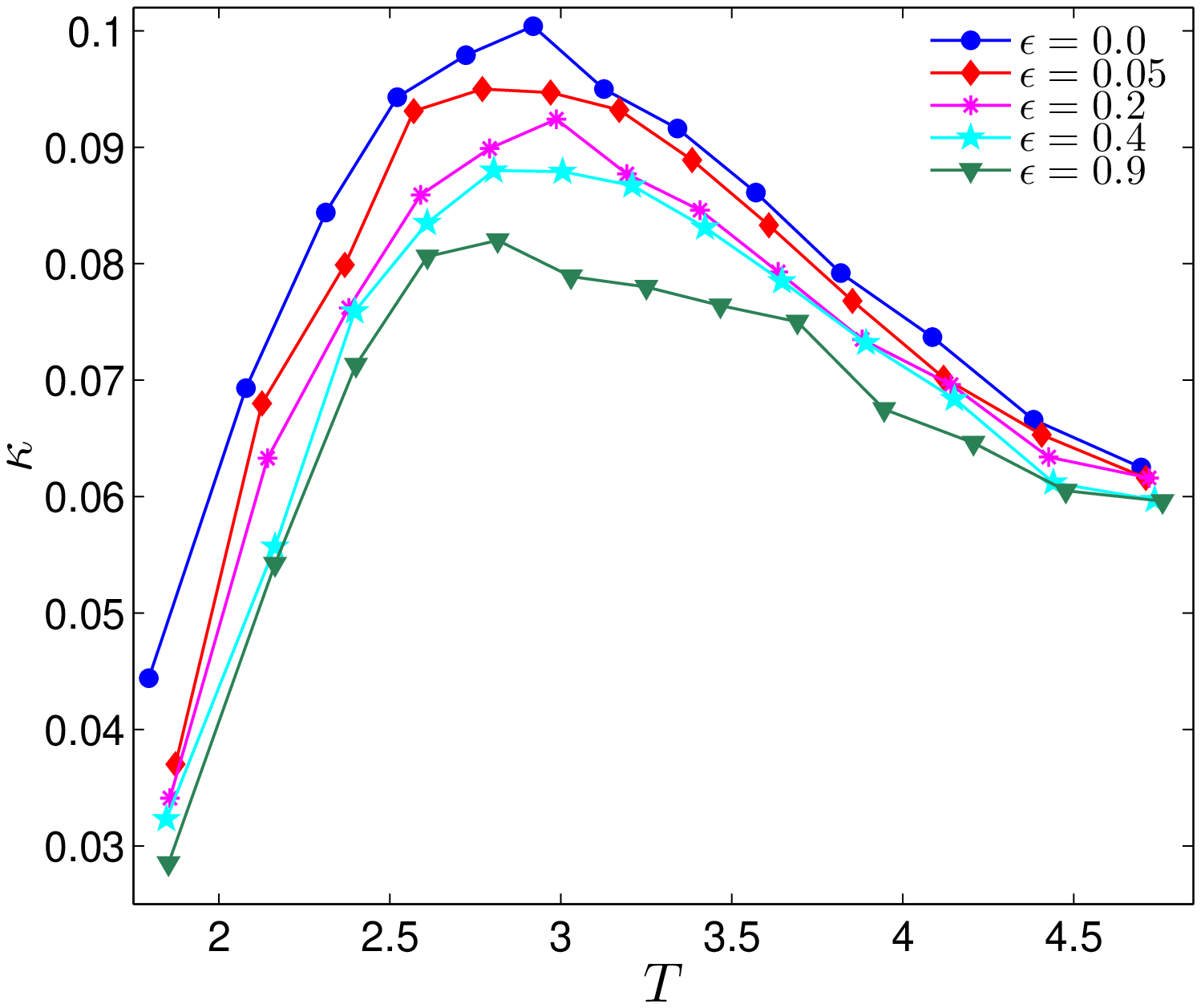}}
\caption{Conductivity versus temperature for a disordered Ising system of size $L = 16$ and fixed temperature gap $T_1=0.01, T_2=5.0$. Different values of disorder $\epsilon$ are considered, as shown in the legend.}

\label{Fig_7}
\end{figure}

\section{Effects of Disorder} \label{sec:effect}

\begin{figure}
\resizebox{0.9\columnwidth}{!}{
\includegraphics{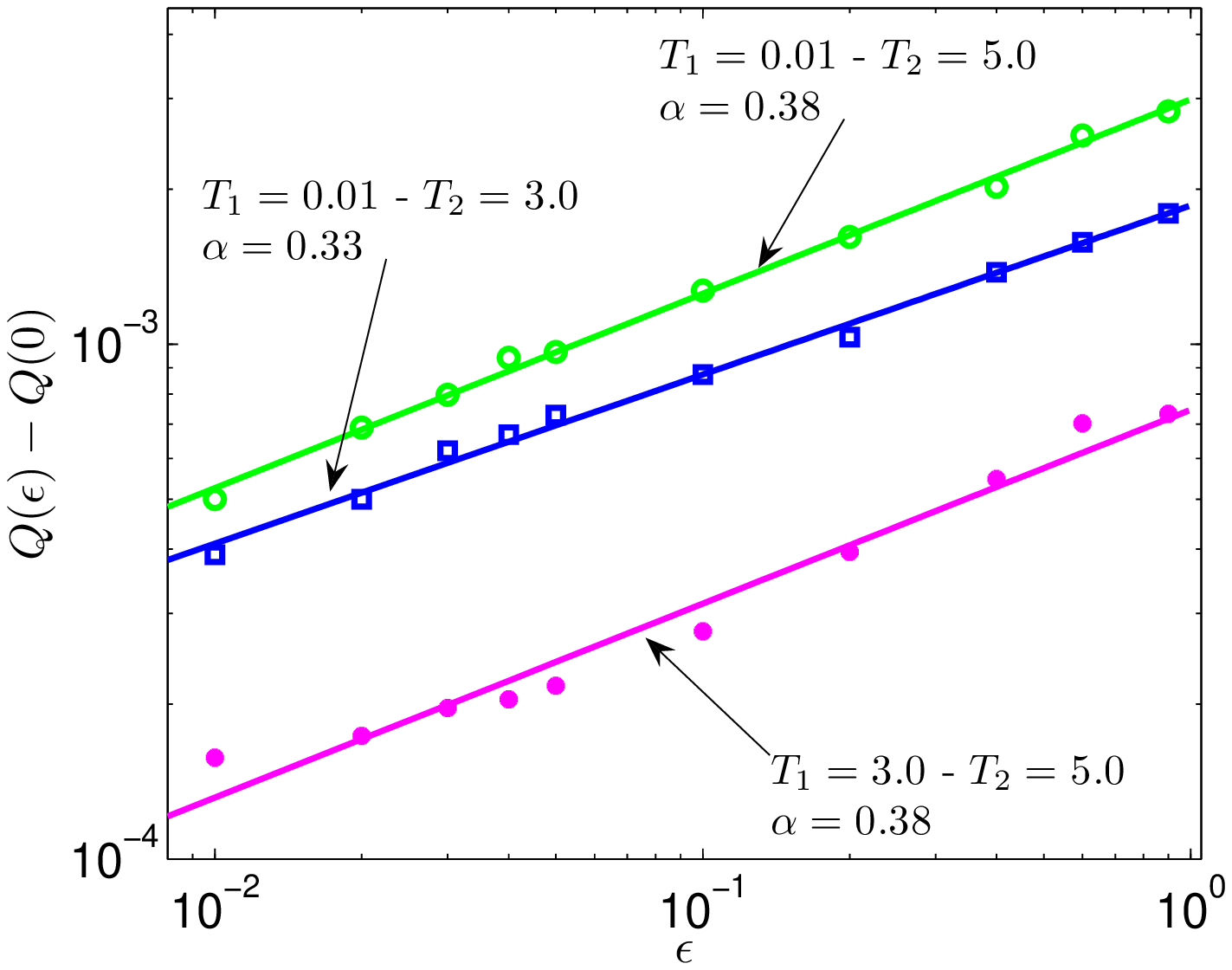}}
\caption{Loglog-scale plot for $Q(0)-Q(\epsilon)$ versus $\epsilon$. Different symbols refer to different temperature gaps, while the continuous line represent the best linear fit. The error on the exponent $\alpha$ is approximately $10 \%$. }
\label{Fig_8}
\end{figure}

Let us now deepen our knowledge of how the conductivity is
affected by disorder $\epsilon$. In Fig.~\ref{Fig_7} we show data
for conductivity vs. temperature for different degrees of
disorder. Remarkably, the figure highlights a clear drift: as the
disorder $\epsilon$ increases the conductivity $\kappa$ gets lower
for any value of the local temperature. Hence, for a given system
characterized by the set of parameters $(L_X,L_Y, T_1, T_2)$ the
heat flow $Q(\epsilon)$ is monotonically decreasing with
$\epsilon$, or, otherwise stated, the presence of disorder lowers
the transport efficiency. We now focus on the behaviour for
$\epsilon \to 0$. The transition from the ordered to the
disordered case is in fact characterized by different expressions
of the kinetic energy as a function of the temperature (see
Eqs.~\ref{emdis2},~\ref{emcon}). This could be a signal of a non
trivial behaviour also for different physical quantities. A
plausible ansatz for a singularity at sufficiently small
$\epsilon$ is provided by the expansion:
\begin{equation}\label{eq:expansion}
Q(\epsilon) \sim Q(0) - A \epsilon^{\alpha},
\end{equation}
where $A>0$ is expected to depend in general on the system parameters (size, gradient $ T_2- T_1$, etc) and $\alpha$ is a possibly universal exponent.
In Fig.~\ref{Fig_8} we show a log-log scale plot for $Q(0)-Q(\epsilon)$ which confirms the scaling law of Eq.~\ref{eq:expansion} over a wide range of disorder and for different choices of temperature intervals.

By means of a fitting procedure it is possible to get estimates for $A$ and $\alpha$. Interestingly, $A$ increases with the temperature gap $T_2-T_1$, and its value is significantly smaller when both temperatures are larger than the critical one, namely $T_2,T_1>T_C$. Hence, as expected, the effect of disorder is especially appreciable for low temperatures. We also notice that, within the error, the agreement among the exponents $\alpha$ is rather good.

Finally, we remark that when $\epsilon \ll 1$ and  both $T_1$ and $T_2$ are smaller than $T_C$ the decorrelation time are long and the measures of $Q(\epsilon)$ and $\alpha$, the relevant exponent, get awkward. Indeed, the ``critical slowing down'' observed for $\epsilon \rightarrow 0$ is neatly distinct from the behaviour at $J=1$ (i.e. $\epsilon=0$). This provides a  hint on a possible non analiticity as $\epsilon \to 0$.

\section{Conclusions} \label{sec:conclusions}
We have introduced a microcanonical dynamics based on the link reserves of kinetic energy, showing that it recovers the results on regular lattices previously obtained by other dynamics and that, moreover, it can be properly applied to disordered systems and it lets a natural definition of temperature, also when far from equilibrium. This allows an easy access to such quantities as conductivity and specific heat, otherwise difficult to define operationally.
We measured the average conductivity for the ordered and the disordered systems showing that the presence of disorder reveals in the reduction of the transport efficiency.

The disordered lattice exhibits remarkable additional features: there are indications for a singular behaviour of the system in the limit $\epsilon  \to 0 $ of the disorder parameter. The reason of such a singularity may lie in the numerical strain in filling the reals by combinations with integer coefficients if the couplings $J_{ij}$ are very close to each other. This slowing down should be the dynamical counterpart of the discontinuity of the temperature (see expressions (\ref{emdis}) and (\ref{emcon})) when the $2\epsilon$-width of the disorder interval tends to zero. Moreover, it would be interesting to highlight the nature of the exponent $\alpha$, which describes how the heat flow decreases with $\epsilon$. In particular, it should be clarified whether $\alpha$ depends on the system parameters $ (L_X,L_Y,T_1,T_2)$ or, if not, whether it depends on the particular dynamics chosen.

Recalling from past experiments that the sole presence of gradients proved influent on the fluctuations of physical and geometrical observables \cite{acv}, on the basis of present results it seems likely that there are other problems deserving future investigations, e.g. the fine interplay between disorder and non-equilibrium in small scale regions, where local heat flows and temperature fluctuations interact presumably in non trivial ways. In other words, for the disordered system we have verified the effectiveness of the Fourier approach by considering averaged quantities, while the validity of such a description at the microscopic scale is still an open problem.

Finally, we underline that  the dynamics introduced here allows possible extensions not only to varied types of disorder (vacancies, negative couplings, dynamical disorder...) or to spin systems lying on different geometrical structures, but also to  systems where the gradient maintaining the condition of non-equilibrium is due, for instance, to the contact with reservoirs at different densities \cite{acv2}. In general, it is conceivable to model some peculiar dynamics where ``conservation'' refers to the sum of two distinct populations.

\appendix
\section{} \label{density}

Let $ \theta_1,...,\theta_m$, with $ m \geq 2$, positive and rationally independent reals. Let $X > 0$.
Then, for arbitrarily small $\delta $, there exists a linear combination with integer coefficients
$$ C = \sum_{k=1}^m N_k \theta_k$$
such that $ 0 \leq  \mid X-C \mid < \delta $.

{\noindent \it Proof~}: let $\lambda _0$ and $M_0$ be the minimal  and the maximal positive reals obtained as integer
combinations $ \sum_{k=1}^m u_k \theta_k$, where $u_k$ may be $1, 0, -1$. Then define $a_0$ as the greatest
positive real $0 < a_0 < \lambda _0$ of the form
$$ a_0 = M_0-k_0 \lambda _0,$$
with integer $k_0$. Such a number $a_0$  is an integer combination of the $ \{ \theta _k \}$
and it does not coincide with $\lambda _0$ because of their rational independence. We define now
$$ \lambda _1 = \min\{ a _0, \lambda _0 -a_0\},$$ observing that necessarily
$\lambda _1 <  \lambda _0/2$.

Now, we iterate the procedure defining
$$ a_n = M_0-k_n \lambda _n,$$
and
$$ \lambda _{n+1} = \min\{a _n, \lambda _n -a_n\}.$$
We obtain a positive sequence $ \lambda _n$ converging to $0$, because
$\lambda _{n+1} <  \lambda _n/2$. Each term in this sequence is an integer combination of
the $ \{ \theta _k \}$. Now, we can choose $N$ in such a way that $\lambda _N < \delta$. Since $\lambda _N $
is finite,  there exists an integer $p$ such that
$$ p ~\lambda _N < X < (p+1)~\lambda _N.$$
We get this way a neighborhood of $X$ smaller than $\delta $, whose extremes are integer combinations of
the $ \{ \theta _k \}$, QED.

\vskip 5.0 pt

\noindent Note 1: the way $X$ has been approximated does not necessarily correspond to an optimal strategy;

\noindent Note 2: with reference to our problem, where the $ \{ \theta _k \}$ are the seven couplings $J_{ij} $ chosen for each link in an interval  $[1-\epsilon,1+\epsilon ]$, and the integer coefficients are the iterated additions and subtractions determined by the wandering spin flips, it seems that, especially for great $X$'s, the probability of the right approximating combination is extremely low. However, in the integrals of formula (\ref{emcon}), the real variable is weighted in turn by the Boltzmann factor, and this explains why also in our finite-time simulations it is possible to to fill densely the relevant part of the real axis, ensuring the correctness of the procedure;

\noindent Note 3: when the $2\epsilon$-width of the interval
shrinks (little disorder), this contributes to make difficult the
approximation: the reduction in numerical efficiency in these
conditions may be observed indeed, as noticed in Section
(\ref{sec:effect}), and it could also be  related to the possible
singularity of the limit $\epsilon \to 0$.

\end{document}